\documentstyle[preprint,tighten,aps,colordvi,psfig,epsfig,rotating]{revtex}

\begin{document}
\draft
\vskip 2cm

\title{\Large{\bf {Additive and multiplicative renormalization of
      topological charge with improved gluon/fermion actions: A test 
      case for 3-loop vacuum calculations, using overlap or clover fermions}}} 

\author{A. Skouroupathis and H. Panagopoulos}
\address{Department of Physics, University of Cyprus, P.O. Box 20537,
Nicosia CY-1678, Cyprus \\
{\it email: }{\tt php4as01@ucy.ac.cy, haris@ucy.ac.cy}}
\vskip 3mm


\maketitle

\begin{abstract}

We calculate perturbative renormalization properties of the topological 
charge, using the standard lattice discretization given by a product of 
twisted plaquettes. We use the overlap and clover action for fermions, and 
the Symanzik improved gluon action for 4- and 6-link loops.
 
We compute the multiplicative renormalization of the topological charge density
to one loop; this involves only the gluon part of the action. The
power divergent 
additive renormalization of the topological susceptibility is
calculated to 3 loops. 

Our work serves also as a test case of the techniques and limitations
of lattice perturbation theory, it being {\bf the first 3-loop computation
in the literature involving overlap fermions.}


\end{abstract}

\newpage

\begin{center}
\Large{\bf Introduction}
\end{center}

Topological properties of QCD are among those most widely studied on
the lattice. Various methods have been used to this end, involving
renormalization, cooling, fermionic zero modes, geometric definitions,
etc.
In recent years, the advent of fermionic actions, such as the overlap,
which do not violate chirality, has brought a new thrust to the subject.

In this work we compute the renormalization constants which are
necessary in order to extract topological properties, in the ``field
theoretic'' approach, from Monte Carlo simulations using {\bf Wilson or
Symanzik improved gluons}, and {\bf clover or overlap fermions}.In
particular, we compute the multiplicative renormalization $\mathbf
Z_Q$ of the topological charge density, to 1 loop in perturbation
theory and the power divergent additive renormalization $\mathbf
M(g^2)$ of the topological susceptibility, to 3 loops.  

The main motivations for doing this work are:

\begin{itemize}
\item[$\bullet$] To enable comparison between different
  approaches used in studying topology, so that a coherent picture of
  topology in QCD may 
  emerge.
\item[$\bullet$] To enable studies, in numerical simulations, of
   quantities involving the density of topological charge, $q(x)$,
   rather than only the integrated charge; this is necessary, e.g., for
   studying the spin content of nucleons.
\item[$\bullet$] As a feasibility study in lattice perturbation
  theory: Indeed, this is the first 3-loop calculation involving
  overlap fermions.
\end{itemize}

\vskip 3mm

\begin{center}
\Large{\bf Computation of $\mathbf Z_{\rm Q}$}
\end{center}

Our first task is to compute the the multiplicative renormalization $Z_Q$
~\cite{CDP} of the topological charge density $q_{_L}(x)$ to one loop, using the background field method.We use the standard definition of $q_{_L}$,
given by a product of twisted plaquettes 
\begin{equation}
q_{_L}(x)=- {1\over 2^9\pi^2}
\sum^{\pm 4}_{\mu\nu\rho\sigma=\pm 1}
\epsilon_{\mu\nu\rho\sigma} {\rm Tr}
\left[ \Pi_{\mu\nu}(x)\Pi_{\rho\sigma}(x)\right]\;\,
\label{stop}
\end{equation}
($\epsilon_{-\mu,\nu,\rho,\sigma}\equiv
-\epsilon_{\mu,\nu,\rho,\sigma}$). $\Pi_{\mu\nu}(x)$ is the parallel transport matrix along a $1 \times{1}$ Wilson loop; in standard notation
\begin{equation}
\Pi_{\mu\nu}(x)=U_\mu(x)\,U_\nu(x+\mu)\,U^\dagger_\mu(x+\nu)\,U^\dagger_\nu(x)
\end{equation}

The classical limit of the operator shown in Eq.(\ref{stop}) must be
corrected by including a renormalization function $Z_Q$, which can be
expressed perturbatively as
\begin{equation}
{\mathbf Z_Q}=1+Z_1\cdot{g^2}+\cdots \,\,,\qquad Z_1={\mathbf Z_{11}}\,
\cdotp N_c+{\mathbf Z_{12}}\,\cdotp{1\over N_c}
\label{ZQ}
\end{equation}
We perform a calculation of $Z_1$; this involves only the gluon part of the action.

In the background field method, link variables are decomposed as
\begin{equation}
U_\mu(x)=V_\mu(x)\,U_{c\mu}(x)
\end{equation}
in terms of links for a quantum field and a classical background
field, respectively
\begin{equation}
V_\mu(x)=e^{igQ_\mu(x)},\qquad
U_{c\mu}(x)=e^{iaB_\mu(x)}
\end{equation}
The $N_c \times{N_c}$ Hermitian matrices $Q_\mu$ and $B_\mu$ can be
expressed as
\begin{equation}
Q_\mu(x)=t^a\,Q_\mu^a(x),\quad B_\mu(x)=t^a\,B_\mu^a(x),\quad Tr
[t^at^b]={1\over 2}\,\delta^{ab} 
\end{equation}

The perturbative nature of our calculation requires a choice of gauge.
An appropriate gauge-fixing term is
\begin{equation}
S_{gf}={1\over {1{-}\xi}}\,\sum_{\mu,\nu}\sum_{x}Tr[D_\mu^{-}Q_\mu
  D_\nu^{-} Q_\nu] 
\end{equation}
This term breaks gauge invariance with respect to $Q_\mu$, as it
should, but succeeds in keeping the path integral as a gauge invariant
functional of $B_\mu$. The definition of the lattice derivative, which
is covariant with respect to background gauge transformations, is
\begin{equation}
D_\mu^{-}(U_c)Q_\nu(x)=U_{c\mu}^{-1}(x-e_\mu)Q_\nu(x-e_\mu)U_{c\mu}(x-e_\mu)-Q_\nu(x)
\end{equation}

The diagrams involved in the one-loop calculation of $Z_Q$ are shown
in Figure 1.

\vspace{1cm}
\begin{center}
\psfig{file=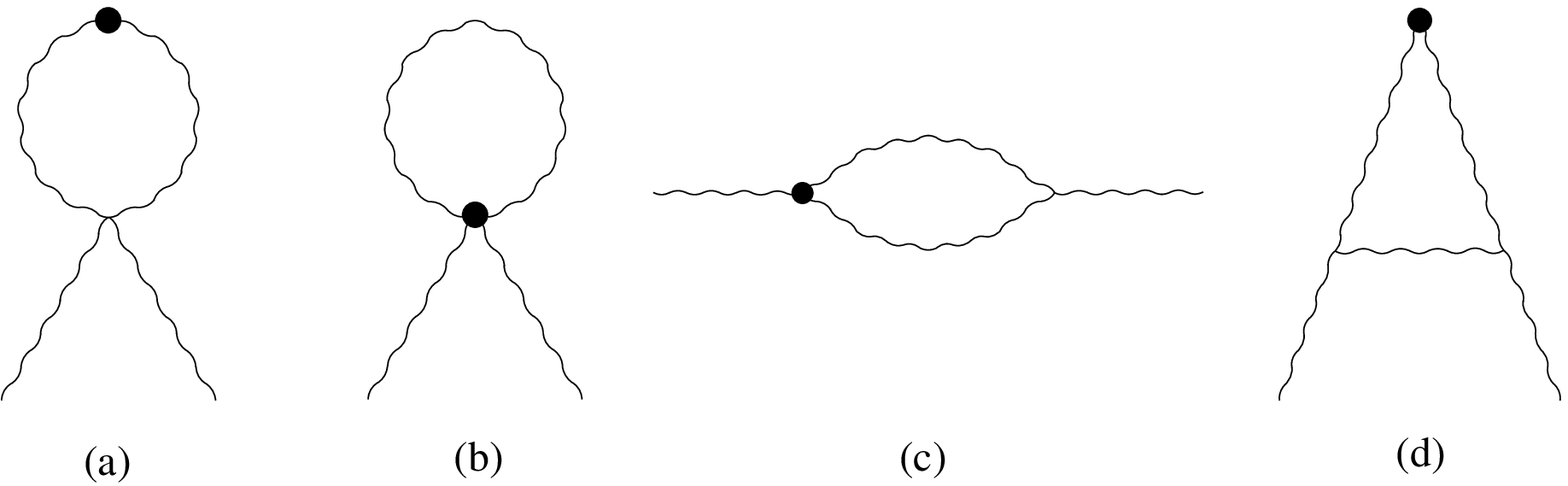,scale=0.7}
\vskip 4mm
\small{\label{fig1}Fig. 1: Diagrams contributing to $Z_1$. The
  black bullet stands for topological charge vertices. External lines
  correspond to background fields.}
\end{center}
\vspace{1cm}

We use the Symanzik improved gauge field action, involving Wilson
loops with 4 and 6 links. In standard notation, it reads
\begin{eqnarray}
S_G=\frac{2}{g^2} \,\,& \Bigg[ &c_0 \sum_{\rm plaquette} {\rm Re\, Tr\,}(1-U_{\rm plaquette})\, \nonumber \\ 
 & + & \, c_1 \sum_{\rm rectangle} {\rm Re \, Tr\,}(1- U_{\rm rectangle}) \nonumber \\ 
 & + & c_2 \,\sum_{\rm chair} {\rm Re\, Tr\,}(1- U_{\rm chair})
\, \nonumber \\ 
 & + & \, c_3 \sum_{\rm parallelogram} {\rm Re \,Tr\,}(1-
U_{\rm parallelogram})\Bigg]\,
\label{gluonaction}        
\end{eqnarray}

The lowest order expansion of this action, leading to the gluon
propagator, is (see, e.g., Ref.~\cite{HPRSS})
\begin{equation}
S_{\rm G}^{(0)} = \frac{1}{2}\int_{-\pi/a}^{\pi/a} \frac{d^4k}{(2\pi)^4}
\sum_{\mu\nu} 
A_\mu^a(k)\left[G_{\mu\nu}(k)-\frac{\xi}{\xi-1}\hat{k}_\mu\hat{k}_\nu\right]
A_\nu^a(-k)\,
\end{equation}
where:\qquad \qquad $G_{\mu\nu}(k) = \hat{k}_\mu\hat{k}_\nu + \sum_\rho \left(
\hat{k}_\rho^2 \delta_{\mu\nu} - \hat{k}_\mu\hat{k}_\rho \delta_{\rho\nu}
\right)  \, d_{\mu\rho}$

\noindent
and:\qquad\qquad $d_{\mu\nu}=\left(1-\delta_{\mu\nu}\right)
\left[C_0 -
C_1 \, a^2 \hat{k}^2 -  C_2 \, a^2( \hat{k}_\mu^2 + \hat{k}_\nu^2)
\right]$
$$ \hat{k}_\mu = \frac{2}{a}\sin\frac{ak_\mu}{2}\,, \quad
        \hat{k}^2 = \sum_\mu \hat{k}_\mu^2 \, $$
The coefficients $C_i$ are related to the Symanzik coefficients $c_i$ by
$$
C_0 = c_0 + 8 c_1 + 16 c_2 + 8 c_3 \,, \,\,\,
C_1 = c_2 + c_3\,, \,\,\, C_2 = c_1 - c_2 - c_3 
$$
The Symanzik coefficients must satisfy: $c_0 + 8 c_1 + 16 c_2 + 8 c_3
= 1$, in order to reach the correct classical continuum limit.

Our calculations are performed without any assumptions on the values
of the external momenta $p_1,\,p_2$ : This is safest for the
topological charge operator, otherwise one may easily end up with
indeterminate expressions. 

For our purpose, we must express all potentially divergent integrals
$I$ in terms of a continuum counterpart $I_{cont}$ (evaluated in
$D=4-2\epsilon$ dimensions), plus all lattice contributions
$I_{Latt}$. The latter are the ones which will determine $Z_Q$. The
three expressions shown below form a basis set for all the divergent
integrals encountered in this calculation; we point out the need to
handle also a 3-point form factor ($C_{\mu\nu}$ below)~\cite{PanVic}
\begin{eqnarray}
\bar{B}(a,p)&=&\frac{(k a)^{2 \epsilon}}{a^0}
\int{{\frac{d^Dk}{(2\pi)^D}} {\frac{1}{\hat{k}^2  \, (\widehat{k+a
	p})^2}}}=B(p)+I_{0_{Latt}} \\ 
\bar{B}_{\mu}(a,p)&=&\frac{(k a)^{2 \epsilon}}{a^1}
\int{{\frac{d^Dk}{(2\pi)^D}} {\frac{sin k_{\mu}}{\hat{k}^2
      \,(\widehat{k+a p})^2}}}=B_{\mu}(p)+I_{1_{Latt}} \\ 
\bar{C}_{\mu\nu}(a,p_1,p_2)&=&\frac{(k a)^{2
    \epsilon}}{a^0}\int{{\frac{d^Dk}{(2\pi)^D}} {\frac{sin k_{\mu} \,
      sin k_{\nu}}{\hat{k}^2 \,(\widehat{k+a p_1})^2(\widehat{k +a p_1
	+a p_2})^2}}} \nonumber \\ 
&=&C_{\mu\nu}(p_1,p_2)+\frac{1}{64
  \pi^2}\delta_{\mu\nu}\bigg(-\frac{1}{\epsilon}-ln \,\kappa^2 a^2 -ln
\, 4\pi \nonumber \\ 
&~& \qquad \qquad \qquad \qquad \quad \,\,+(4\pi)^2\,P_2+\gamma_E -2
\pi^2 P_1\bigg) 
\end{eqnarray}
where $\gamma_E$ is Euler's constant,
$B,\,B_{\mu},\,C_{\mu\nu}$ are the continuum
counterparts of
$\bar{B},\,\bar{B}_{\mu},\,\bar{C}_{\mu\nu}$, 
respectively; $P_1,\,P_2$ are~\cite{LW}
$$ P_1=0.15493339023109021(1), \quad P_2=0.02401318111946489(1) $$
\begin{eqnarray}
{\rm and:} \qquad I_{0_{Latt}}&=&\frac{1}{(4
  \pi)^2}\Big(-\frac{1}{\epsilon}-ln \,4\pi -ln \,\kappa^2 a^2 +
\gamma_E \Big)+ P_2  \\
I_{1_{Latt}}&=&p_{\mu}\Bigg(\frac{P_1}{16}-{1\over2}
  I_{0_{Latt}}\Bigg) \nonumber  
\end{eqnarray}
As we see, $I_{Latt}$ contains poles in $\epsilon$;
indeed, diagrams (c) and (d) of Figure 1, taken separately, exhibit such
poles ( (d) $\propto -1/\epsilon-ln \,\kappa^2
a^2$). These cancel, however, upon summation, as is expected by the
fact that $Q$ does not renormalize in the continuum. 

The complicated algebra of perturbation theory was carried out using our
package in Mathematica.
The calculation of $Z_Q$ is particularly involved in
the present case, involving propagators and vertices from the improved
gluonic action. In particular, the calculation of diagram (d) involves a
summation of {\bf $> 1\,000\,000$ different algebraic expressions} at
intermediate stages.

Our results for $Z_Q$ are listed in Table I.
In all calculations that involve the parameters $c_i$, we choose a standard set of values as in Ref.~\cite{HPRSS}. The choice of the sets of parameters correspond to the most popular actions:
The first set corresponds to the plaquette action, the second set
corresponds to the tree-level Symanzik improved action~\cite{Symanzik} 
and the next 6
sets correspond to the tadpole improved L\"uscher-Weisz (TILW)
action~\cite{Luscher,Alford}
for 6 values of beta 
\begin{center}
$\beta =8.60,\,8.45,\,8.30,\,8.20,\,8.10,\,8.00$
\end{center}
The last two sets correspond to the Iwasaki~\cite{Iwasaki} and
 DBW2~\cite{Takaishi} actions
 respectively.

In the case of the plaquette action, our result
agrees with the known result of Ref.~\cite{CDP}.

It is worth noting that the value of $Z_1$ (and of
$e_3$ in Table II) for the DBW2 action is the smallest one, leading to
a renormalization factor $Z_Q$ closer to 1 (and $M(g^2)$ closer to
0, see below. {\bf This would single out the DBW2 action as a better
  candidate for 
studies of topology}.

\vspace{0.4cm}

\begin{center}
\Large{\bf Computation of $\mathbf M(g^2)$}
\end{center}

The second task we attend to is the calculation of the power additive renormalization of the topological charge susceptibility, which is defined as
\begin{equation}
\chi_{_L}\;=\; \sum_x \langle q_{_L}(x)q_{_L}(0) \rangle 
\label{childef}
\end{equation}
$\chi_{_L}$ develops an unphysical background term which becomes
dominant in the continuum limit
\begin{equation}
\chi_{_L}(g^2)\;=\;a^4 Z_Q(g^2)^2\chi\,+\,M(g^2)
\label{chileq}
\end{equation}
The power divergent additive renormalization of $\chi_{_L}$ can be written perturbatively as
\begin{equation}
M(g^2)={\mathbf e_3}\cdot{g^6}+{\mathbf e_4}\cdot{g^8}+\cdots
\label{Mg2}
\end{equation}
We first compute the 2-loop coefficient $e_3$\,. Its dependence on the
number of colors has the form:
\begin{equation}
{\mathbf e_3} = (N_c^2-1)\,N_c\,{\mathbf e_{3,0}}
\label{e3}
\end{equation}
This result is evaluated for several
sets of values of the Symanzik improvement coefficients. Figure 2 shows
the diagram contributing to $e_3$\,. 
\vspace{1cm}
\begin{center}
\psfig{file=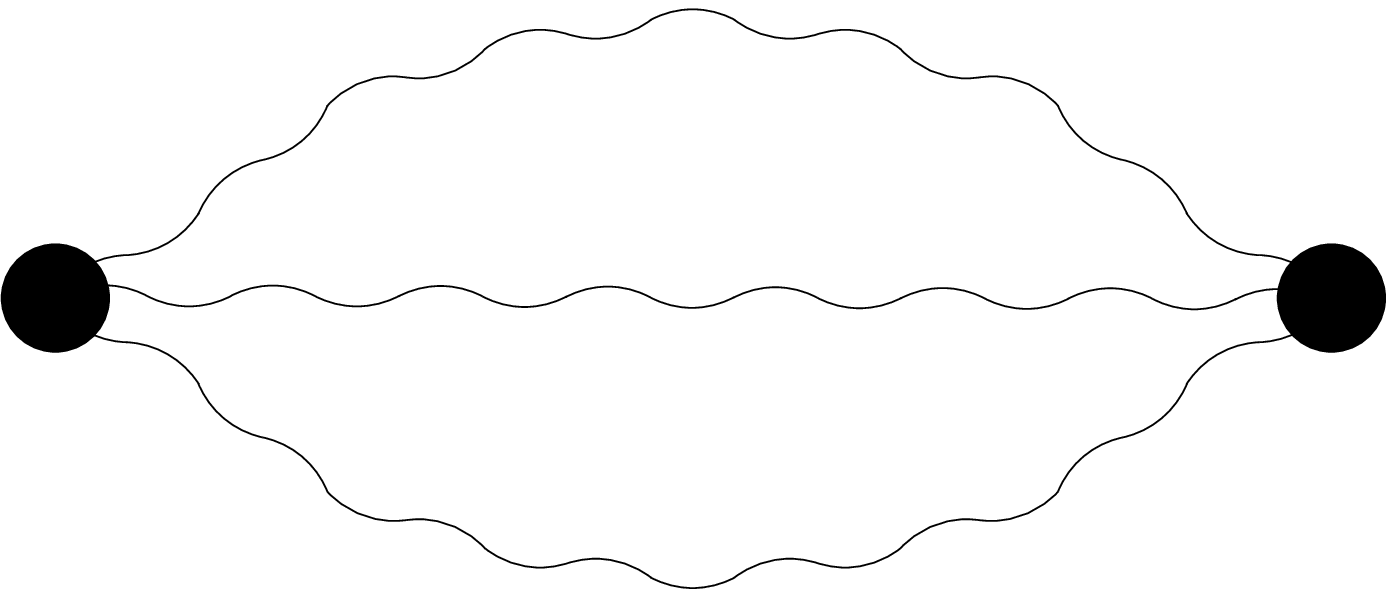,scale=0.25}
\vskip 4mm
\small{\label{fig2}Fig. 2: Two-loop diagram contributing to
  ${e_3}$. Black bullets stand for topological charge vertices.}
\end{center}

 The 3-loop term $e_4$ of the expansion of $M(g^2)$ equals
\begin{equation}
e_4=e_{4}^g+{\mathbf e_{4}^f}
\label{e4}
\end{equation}
where $e_{4}^f$ stands for the fermionic contribution to $e_4$ 
($c_{\rm SW}$ is the coefficient in the clover action) 
\begin{equation}
{\mathbf e_{4}^f}=N_f(N_{c}^2-1)N_c
\cdot({\mathbf e_{4,0}}+{\mathbf e_{4,1}}\,c_{\rm
  SW}+{\mathbf e_{4,2}}\,c_{\rm SW}^2) 
\label{e4f}
\end{equation}
and $e_{4}^g$
is the purely gluonic contribution, which is expressed as in Ref.~\cite{ACFP}
\begin{equation}
e_{4}^g={1\over{16}}(N_{c}^2-1)(1.735N_{c}^2-10.82+73.83/N_{c}^2) \times{10^{-7}}
\end{equation}
In fact, what we are interested in, is the calculation of the
parameters $e_{4,0},\,e_{4,1},\,e_{4,2}$. {\bf This task is performed
using both overlap and clover fermions} (Clearly, overlap fermions
involve only the parameter $e_{4,0}$\,). Figure 3 shows the 3-loop
diagrams contributing to the evaluation of $e_4^f$. 

\vskip 5mm
\begin{center}
\psfig{file=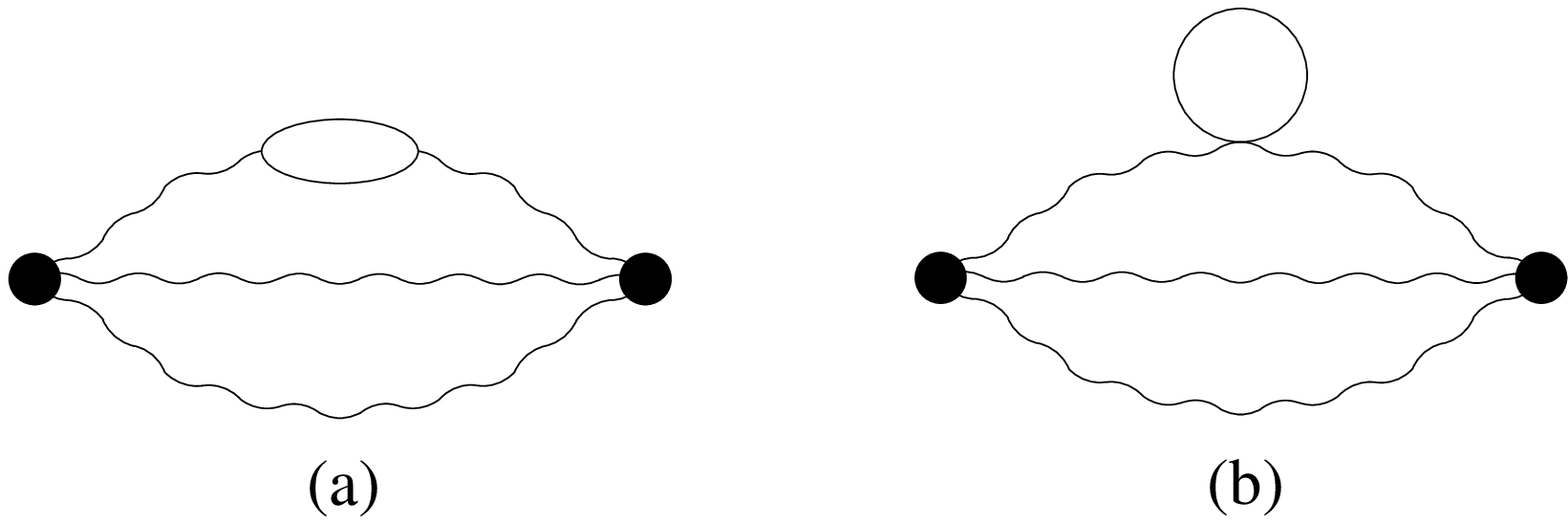,angle=90,scale=0.5}
\vskip 2mm
\small{\label{fig3}Fig. 3: Diagrams contributing to $e_4^f$. Straight lines correspond to \\ overlap or clover fermions and wavy lines correspond to Wilson gluons.}
\end{center}

The propagator and vertices for overlap fermions can
be obtained from the following expression of the overlap action
\begin{equation}
S_{Overlap}=a^4\sum_{n,\,m}\mathit{\bar{\Psi}(n)}D_N(n,m)\mathit{\Psi(m)}
\end{equation}
where $D_N$ is the massless Neuberger-Dirac operator~\cite{Neuberger}
\begin{equation}
D_N = \frac{M_0}{a} \left( 1+ \frac{X}{\sqrt{X^\dagger X}}\right)
\label{over}
\end{equation}
$M_0$ is a real parameter corresponding to a negative mass term. $M_0$
must lie in the range $0 < M_0 < 2r$, r being the Wilson parameter (in
our case $r=1$). $X$ is a Hermitian operator which can be expressed,
in momentum space, in terms of the Wilson-Dirac operator $D_W$ 
\begin{eqnarray}
X(q,p)&=&a \,\left(D_{\rm W}(q,p) - {M_0\over a}\right) \\ 
&=&a X_0(p)(2\pi)^4\delta^4(q-p)+a X_1(q,p)+aX_2(q,p) + {\cal O}(g^3)\nonumber
\end{eqnarray}

\begin{eqnarray}
{\rm where:\quad} X_0(p)&=&{i\over a} \sum_\mu \gamma_\mu \sin a p_\mu + {r\over a} \sum_\mu (1-\cos a p_\mu) - {M_0\over a}\\
X_1(q,p)&=&g\int d^4 k\, \delta(q-p-k) A_\mu(k) V_{1,\mu}(p+{k\over 2})\\
{\rm with:\quad} V_{1,\mu}(q)&=&i\gamma_\mu \cos aq_\mu + r\sin a
q_\mu\nonumber\\ 
{\rm and:\quad} X_2(q,p)&=&{g^2\over 2} \int {d^4 k_1 \, d^4 k_2\over (2\pi)^4}
 \delta(q-p-k_1-k_2)\\
&&\qquad A_\mu(k_1)A_\mu(k_2) V_{2,\mu}(p+{k_1\over 2}+{k_2\over 2})\nonumber\\
{\rm with:\quad} V_{2,\mu}(q)&=&-i\gamma_\mu a\sin aq_\mu + ra \cos a
q_\mu\nonumber 
\end{eqnarray}

The clover (SW) fermionic action~\cite{SW}, in
standard notation, reads 
\begin{eqnarray}
S_L &=& {1\over g^2} \sum_{x,\,\mu,\,\nu}
{\rm Tr}\left[ 1 - \Pi_{\mu\nu}(x) \right]  +
\sum_{f}\sum_{x} (4r+m)\bar{\psi}_{f}(x)\psi_f(x) 
\nonumber \\
&-& {1\over 2}\sum_{f}\sum_{x,\,\mu}
\bigg{[}\bar{\psi}_{f}(x) \left( r - \gamma_\mu\right)
U_{\mu}(x)\psi_f(x+{\mu}) \nonumber \\
&~& \hspace{1.7cm}+\bar{\psi}_f(x+{\mu})\left( r + \gamma_\mu\right)U_{\mu}(x)^\dagger\psi_{f}(x)\bigg{]}\nonumber \\
&+& {i\over 4}\,c_{\rm SW}\,\sum_{f}\sum_{x,\,\mu,\,\nu} \bar{\psi}_{f}(x)
\sigma_{\mu\nu} {\hat F}_{\mu\nu}(x) \psi_f(x),
\label{latact}
\end{eqnarray}
\begin{eqnarray}
{\rm where:}\qquad {\hat F}_{\mu\nu} &\equiv& {1\over{8a^2}}\,
(Q_{\mu\nu} - Q_{\nu\mu})\\
{\rm and:\qquad} Q_{\mu\nu} &=& U_{x,\, x+\mu}U_{x+\mu,\, x+\mu+\nu}U_{x+\mu+\nu,\, x+\nu}U_{x+\nu,\, x}\nonumber \\
&+& U_{ x,\, x+ \nu}U_{ x+ \nu,\, x+ \nu- \mu}U_{ x+ \nu- \mu,\, x- \mu}U_{ x- \mu,\, x} \nonumber \\
&+& U_{ x,\, x- \mu}U_{ x- \mu,\, x- \mu- \nu}U_{ x- \mu- \nu,\, x- \nu}U_{ x- \nu,\, x}\nonumber \\
&+& U_{ x,\, x- \nu}U_{ x- \nu,\, x- \nu+ \mu}U_{ x- \nu+ \mu,\, x+ \mu}U_{ x+ \mu,\, x}
\label{latact2}
\end{eqnarray}

The clover coefficient {\bf $\mathbf c_{\rm SW}$ is treated here as a free parameter};
$r$ is the Wilson parameter;
$f$ is a flavor index; $\sigma_{\mu\nu} =(i/2) [\gamma_\mu,\,\gamma_\nu]$. 
Powers of the lattice spacing $a$ have been omitted and may be
directly reinserted by dimensional counting. 

In performing this calculation, a large effort was
devoted to the {\bf creation  of an efficient 3-loop ``integrator''}, that
is, a metacode for converting lengthy 3-loop integrands into efficient
code for numerical integration. Some key features of the integrator
are listed in the Appendix.

Table II contains our results for $e_3$
(cf. Eqs.(\ref{Mg2}, \ref{e3})) for different gluonic actions. Fermions do not
contribute here. Our results for the case of the plaquette action
agree with older known results (see, e.g., \cite{diVecchia,Christou}). Tables
III and IV list our results for $e_4$, using the clover and overlap
actions, respectively.

Figure 4 shows the coefficients $e_{4,0},\ e_{4,1},\
e_{4,2}$ of the clover result for different values of the bare fermion
mass $m$. For ease of
reference, Figure 5 presents the total values of $e_4$ for different
choices of $c_{\rm SW}$, with $N_f=2,\ N_c=3$.

Figure 6 exhibits the dependence of $e_{4,0}$ , using
the overlap action, on the parameter $M_0$. The total value of $e_4$
in this case is shown in Figure 7.

\vspace{0.5cm}
\begin{center}
\Large{\bf Discussion and Conclusions}
\end{center}

We have calculated in this work the perturbative renormalization
coefficients for the topological charge and susceptibility. The
topological charge, defined via a usual ``twisted'' product of
plaquettes, was renormalized multiplicatively to one loop using a
variety of gluon actions presently employed in numerical
simulations. The details of the fermionic action only enter this
computation at 2-loop order. The 1-loop coefficient is rather
pronounced for most actions but it is quite suppressed for the
particular case of the DBW2 action.

For the topological susceptibility, we performed a calculation of its
power-divergent, additive renormalization $M(g^2)$ to 2 and 3
loops. Fermions do not contribute to the 2-loop result. Among the
various gluonic actions, the DBW2 action leads again to a coefficient
which is quite suppressed, by more than 2 orders of magnitude as
compared to the standard Wilson action. Based on these results, the
DBW2 action is singled out as the favourite candidate for simulations
involving measurements of topology in the ``field theoretic'' approach.

The 3-loop calculation was performed in the presence of clover, as
well as overlap fermions, and Wilson gluons. Indeed, this is the first
3-loop calculation
in the literature involving overlap fermions. To facilitate a
comparison of various contributions to $M(g^2)$, we write them in the
particular case of $N_c=3$ with Wilson gluons:
\begin{equation}
M(g^2) = g^6 \, 1.655\times 10^{-5} + g^8 \, 6.50 \times 10^{-7}
+ 24\,g^8 \, N_f\, x
\end{equation}

The first two summands above are clearly the purely gluonic
contributions, at 2 and 3 loops. The value of $x$ can be read from
Table III for clover fermions ($x = e_{4,0} + e_{4,1}\, c_{\rm SW} +
e_{4,2}\, c_{\rm SW}^2$) and from Table IV for overlap fermions
($x=e_{4,0}$). In all cases, fermionic contributions are of the same
order of magnitude as 3-loop gluon contributions, but
still only a small fraction of the 2-loop result.
 
Our results can be used to enhance a number of
related computations. In some cases, the local definition of
$q_{_L}(x)$ is used (e.g.,~\cite{ADG}), with renormalization
estimates coming from simulations; other investigations propose
non-ultralocal definitions of 
$q(x)$ (e.g.,~\cite{GRT}), which clearly are more expensive
to simulate, but can circumvent renormalization; in yet other cases,
operational/numerical definitions of $q(x)$ are proposed
(e.g.,~\cite{MILC}). Even for the integrated topological charge $Q$,
where fermions with exact chiral symmetry offer us in principle
unambiguous ways of extracting $Q$ from lattice configurations (see,
e.g., Refs.~\cite{Giusti,DelDebbio}), a comparison with other
time-tested definitions is called for. In all cases, it would be very 
important to verify that a consistent 
picture of topology in QCD emerges from the various approaches.

\vspace{0.5cm}
\noindent
{\Large{\bf Appendix}}

\bigskip
Here we briefly describe some features of our ``integrator''
program; this is metacode, written in Mathematica, for converting
lengthy 3-loop integrands into efficient Fortran
code for numerical integration on finite lattices of different sizes $L^4$.
The results $r(L)$ are subsequently extrapolated to $L\to\infty$, via a fit
to a large class of functions containing powers of $L$ and $\ln L$,
as is {\it a priori} expected; systematic errors on $r(\infty)$ are
also produced as a result of these fits.

A number of optimizations are implemented, 
as listed below; the resulting output code runs many
orders of magnitude faster than output of automatic code
conversion programs, such as, e.g., the built-in converter of Mathematica.

The input expression to the integrator is an integrand which
depends on 3 four-momenta $p_1,\ p_2,\ p_3$\,. It has the form of a
sum of terms (typically tens of thousands); each term is a product
of trigonometric functions of combinations of $p_i$\,, possibly
including also propagators for Symanzik improved gluons and overlap
fermions. 

\smallskip
The following optimizations take place:

\noindent
\hspace{-0.5cm} $\bullet$ 
A number of time consuming common ingredients are precalculated
  numerically and stored, for all possible momentum values on a finite
  lattice; such ingredients are trigonometric functions, the overlap
  propagator, and the Symanzik propagator. While an
  expression in closed form exists for the latter \cite{HPRSS}, in practice
  it is considerably faster to invert the propagator numerically and
  store it; after all, in either case the final results cannot be
  presented as analytic functions of the parameters $c_i$\,, since the
  dependence on $c_i$ is not polynomial (unlike the case of $c_{\rm SW}$).

\noindent
\hspace{-0.5cm} $\bullet$ 
The symmetries of the integrand are exploited to reduce
  the volume of the integration region, with due attention paid to
  correct counting of points at the borders.

\noindent
\hspace{-0.5cm} $\bullet$ 
All 3-loop diagrams, with the exception of those in the form of
  the Mercedes emblem, contain two non-overlapping loops (i.e., loops
  with no shared propagators); consequently, their integrands can
  always be written (after some trigonometry) as a sum over expressions of
  the form:
\begin{equation}
\int dp^4_1 f_1(p_1)\,\Bigl(\int dp^4_2 f_2(p_1,p_2)\Bigr)\, 
                  \Bigl(\int dp^4_3 f_3(p_1,p_3)\Bigr)
\end{equation}
Integrations over $p_2$ and $p_3$ can then be performed sequentially,
rather than in a nested fashion (actually, we perform them
simultaneously, see next item), thus reducing their execution time to
that of a 2-loop integral.

\noindent
\hspace{-0.5cm} $\bullet$ 
The integrand is organized as an inverse tree, i.e. summands
  having the same functional dependence on the innermost integration
  variable are grouped together, and the innermost integral is
  performed once for each group; the procedure is then iterated for
  outer integrals. In practice, this procedure can save one or two
  orders of magnitude in execution time of the innermost
  integrals, which are the most expensive ones.

\noindent
\hspace{-0.5cm} $\bullet$ 
Terms with different polynomial dependence on unspecified parameters such
  as $N_c$ or $c_{\rm SW}$, etc.,
  but with otherwise similar functional form are treated
  simultaneously, and the result is presented as a polynomial in these
  parameters. Also, for parameters whose values are read on input
  (masses, Symanzik coefficients), the code runs in parallel for
  different sets of values, in order to avoid computing the same
  quantities several times.


\begin{table}[ht]
\begin{center}
\begin{minipage}{16cm}
\caption{The values of $\mathbf Z_{11}$ and $\mathbf Z_{12}$ 
(Eq.(\ref{ZQ}), Figure 1), for various values of the
  Symanzik coefficients $c_0,\,c_1,\,c_2,\,c_3 \ (c_2=0)$. 
\label{tab1}}
\bigskip
\begin{tabular}{lllr@{}lrr@{}lllr@{}lllr@{}llllr@{}l}
\multicolumn{1}{c}{Action}&
\multicolumn{4}{c}{$c_0$}&
\multicolumn{5}{c}{$c_1$}&
\multicolumn{2}{c}{$c_3$}&
\multicolumn{4}{c}{$Z_{11}$}&
\multicolumn{5}{c}{$Z_{12}$} \\
\tableline \hline
Plaquette &&&  1&.0  && 0&.0 &&&  0&.0 &&&   -0&.33059398205(2) &&&&  0&.2500000000(1) \\
Symanzik  &&&   1&.6666666 && -0&.083333 &&&  0&.0 &&& -0&.2512236240(1) &&&&  0&.183131339233(1) \\
TILW, $\beta=8.60$  &&&   2&.3168064  && -0&.151791 &&& -0&.0128098 &&&  -0&.20828371039(3) &&&&  0&.147519438874(3) \\
TILW, $\beta=8.45$  &&&   2&.3460240  && -0&.154846 &&& -0&.0134070 &&&  -0&.20674100461(1) &&&&  0&.146259768983(1) \\
TILW, $\beta=8.30$  &&&   2&.3869776  && -0&.159128 &&& -0&.0142442 &&&  -0&.20462181183(1) &&&&  0&.144531861677(4) \\
TILW, $\beta=8.20$  &&&   2&.4127840  && -0&.161827 &&& -0&.0147710 &&&  -0&.20331145580(1) &&&&  0&.143464931830(1) \\
TILW, $\beta=8.10$  &&&   2&.4465400  && -0&.165353 &&& -0&.0154645 &&&  -0&.20162651307(1) &&&&  0&.142094444611(2) \\
TILW, $\beta=8.00$  &&&   2&.4891712  && -0&.169805 &&& -0&.0163414 &&&  -0&.19954339172(1) &&&&  0&.140402610424(1) \\
Iwasaki  &&&   3&.648 && -0&.331 &&& 0&.0 &&&  -0&.15392854668(1) &&&&  0&.105132852383(2) \\
DBW2  &&&   12&.2688  && -1&.4086 &&& 0&.0 &&&  -0&.0617777059(4) &&&&  0&.038277296152(6) \\
\end{tabular}
\end{minipage}
\end{center}
\end{table}

\begin{table}[ht]
\begin{center}
\begin{minipage}{14cm}
\caption{Evaluation of $\mathbf e_{3,0}$ (cf. Eqs.(\ref{Mg2},
  \ref{e3}), Figure 2)
  with Symanzik improved gluons, for various values of the
  coefficients $c_0,\,c_1,\,c_2,\,c_3 \ (c_2=0)$.  
\label{tab2}}\bigskip
\begin{tabular}{lllr@{}llrrr@{}llllr@{}lllr@{}l}
\multicolumn{1}{c}{Action}&
\multicolumn{5}{c}{$c_0$}&
\multicolumn{7}{c}{$c_1$}&
\multicolumn{4}{c}{$c_3$}&
\multicolumn{2}{c}{$e_{3,0}\times{10^{7}}$} \\
\tableline \hline 
Plaquette &&&  1&.0  &&&& 0&.0 &&&&  0&.0 &&&   6&.89791329(1) \\
Symanzik &&& 1&.6666666 &&&& -0&.083333 &&&&  0&.0 &&& 3&.1814562840(7) \\
TILW, $\beta=8.60$  &&&   2&.3168064  &&&& -0&.151791 &&&& -0&.0128098 &&&  1&.8452250005(2) \\
TILW, $\beta=8.45$  &&&   2&.3460240  &&&& -0&.154846 &&&& -0&.0134070 &&&  1&.8054229585(4) \\
TILW, $\beta=8.30$  &&&   2&.3869776  &&&& -0&.159128 &&&& -0&.0142442 &&&  1&.7516351593(8) \\
TILW, $\beta=8.20$  &&&   2&.4127840  &&&& -0&.161827 &&&& -0&.0147710 &&&  1&.7188880608(5) \\
TILW, $\beta=8.10$  &&&   2&.4465400  &&&& -0&.165353 &&&& -0&.0154645 &&&  1&.6773505020(9) \\
TILW, $\beta=8.00$  &&&   2&.4891712  &&&& -0&.169805 &&&& -0&.0163414 &&&  1&.626880218(1) \\
Iwasaki  &&&   3&.648 &&&& -0&.331 &&&& 0&.0 &&&  0&.752432061(7) \\
DBW2  &&&   12&.2688  &&&& -1&.4086 &&&& 0&.0 &&&  0&.04881939(4) \\
\end{tabular}
\end{minipage}
\end{center}
\end{table}

\linespread{1}
\begin{table}[ht]
\begin{center}
\begin{minipage}{14cm}
\caption{Evaluation of {$\mathbf e_4^f$} (cf. Eqs.(\ref{Mg2}, \ref{e4},
  \ref{e4f}), Figure 3), with Wilson gluons and clover fermions, for
  various values of the bare fermion mass $m$.  
\label{tab3}}
\bigskip
\begin{tabular}{r@{}lllr@{}lrrrr@{}lllr@{}lr@{}l}
\multicolumn{4}{c}{$m$}&
\multicolumn{2}{c}{$e_{4,0}\times{10^{8}}$}&
\multicolumn{7}{c}{$e_{4,1}\times{10^{8}}$}&
\multicolumn{2}{c}{$e_{4,2}\times{10^{8}}$} \\
\tableline \hline
-1&.0149250  &&&  -4&.6273(2)  &&&&& 1.28551(1) &&& &   -1.85010(5) \\
-0&.9512196  &&& -4&.3888(2)  &&&&& 1.17807(9) &&& & -1.83818(4) \\
-0&.8749999  &&&   -4&.1089(2)  &&&&& 1.05421(9) &&& &  -1.82249(3) \\
-0&.8253968  &&&   -3&.9299(2)  &&&&& 0.97650(7) &&&  & -1.81156(1) \\
-0&.7948719  &&&   -3&.8210(2)  &&&&& 0.92984(5) &&& &  -1.80457(1) \\
-0&.5181059  &&&   -2&.8759(3)  &&&&& 0.54981(2) &&& &  -1.73390(5) \\
-0&.4234620  &&&   -2&.57089(6)  &&&&& 0.43874(5) &&& &  -1.70723(3) \\
-0&.4157708  &&&   -2&.54658(2)  &&&&& 0.43016(6) &&& &  -1.70500(2) \\
-0&.4028777  &&&   -2&.5061(2)  &&&&& 0.41594(7) &&& &  -1.70128(3) \\
-0&.3140433  &&&   -2&.2325(4)  &&&&& 0.32325(7) &&& &  -1.67496(2) \\
-0&.3099631  &&&   -2&.2202(4)  &&&&& 0.31925(9) &&& &  -1.67373(1) \\
-0&.3017750  &&&   -2&.1956(4)  &&&&& 0.31127(7) &&& &  -1.67125(1) \\
-0&.2962964  &&&   -2&.1793(3)  &&&&& 0.30597(2) &&& &  -1.66958(1) \\
-0&.2852897  &&&   -2&.1460(4)  &&&&& 0.2953(2) &&& &  -1.66621(2) \\
-0&.2825278  &&&   -2&.1379(4)  &&&&& 0.2927(2) &&& &  -1.66536(2) \\
-0&.2769916  &&&   -2&.1215(5)  &&&&& 0.2876(1) &&& &  -1.66366(2) \\
-0&.2686568  &&&   -2&.0972(1)  &&&&& 0.2798(2) &&& &  -1.66108(3) \\
-0&.1482168  &&&   -1&.7519(2)  &&&&& 0.17636(9) &&& &  -1.62264(9) \\
0&.0000  &&&   -1&.36897(4)  &&&&& 0.077477(3) &&& &  -1.57092(3) \\
0&.0050  &&&   -1&.35710(1)  &&&&& 0.074754(4) &&& &  -1.56906(3) \\
0&.0100  &&&  -1&.34534(4)  &&&&& 0.072076(3) &&& &   -1.56720(3) \\
0&.0140  &&& -1&.33599(4)  &&&&& 0.069967(1) &&& & -1.56570(2) \\
0&.0160  &&&   -1&.33134(4)  &&&&& 0.068929(3) &&& &  -1.56494(2) \\
0&.0180  &&&  -1&.32671(4)  &&&&& 0.067895(2) &&& &  -1.56419(1) \\
0&.0236  &&&   -1&.31382(4)  &&&&& 0.065038(5) &&& &  -1.56207(1) \\
0&.0270  &&&   -1&.30603(5)  &&&&& 0.063335(4) &&& &  -1.56077(3) \\
0&.0350  &&&   -1&.28801(6)  &&&&& 0.09418(3) &&& &  -1.55773(6) \\
0&.0366  &&&   -1&.28443(5)  &&&&& 0.058649(4) &&& &  -1.55711(6) \\
0&.0380  &&&   -1&.28131(5)  &&&&& 0.057982(4) &&& &  -1.55658(6) \\
0&.0427  &&&   -1&.27089(5)  &&&&& 0.055767(4) &&& &  -1.55476(7) \\
0&.0460  &&&   -1&.26363(5)  &&&&& 0.054238(5) &&& &  -1.55347(7) \\
0&.0535  &&&   -1&.24729(4)  &&&&& 0.05084(2) &&& &  -1.55051(5) \\
0&.0550  &&&   -1&.24405(5)  &&&&& 0.05018(2) &&& &  -1.54991(5) \\
0&.0720  &&&   -1&.20807(2)  &&&&& 0.04295(2) &&& &  -1.54317(6) \\
0&.0927  &&&   -1&.1658(2)  &&&&& 0.03486(3) &&& &  -1.53475(5) \\
\end{tabular}
\end{minipage}
\end{center}
\end{table}

\begin{table}[ht]
\begin{center}
\begin{minipage}{10cm}
\caption{Evaluation of {$\mathbf e_{4,0}$}
  (cf. Eqs.(\ref{Mg2}, \ref{e4}, \ref{e4f}), Figure 3), with Wilson
  gluons and overlap fermions, for various values of $M_0$,\ \  $0<M_0<2$.
\label{tab4}}
\bigskip
\begin{center}
\begin{minipage}{6cm}
\begin{tabular}{r@{}lr@{}l}
\multicolumn{2}{c}{$M_0$}&
\multicolumn{2}{c}{$e_{4,0}\times{10^{8}}$} \\
\tableline \hline
0&.01   &  0&.59855(5) \\
0&.05   &  0&.63347(4) \\
0&.10    &  0&.6769(2) \\
0&.20    &  0&.7628(2) \\
0&.30    &  0&.8451(2) \\
0&.40    &  0&.92220(3) \\
0&.50    &  0&.99357(4) \\
0&.60    &  1&.05872(3) \\
0&.70    &  1&.11725(2) \\
0&.80    &  1&.16893(3) \\
0&.90    &  1&.213650(1) \\
1&.00    &  1&.251396(7) \\
1&.10    &  1&.282246(6) \\
1&.20    &  1&.306376(2) \\
1&.30    &  1&.32406(2) \\
1&.40    &  1&.33561(4) \\
1&.50    &  1&.34149(2) \\
1&.60    &  1&.34201(7) \\
1&.70    &  1&.3373(2) \\
1&.80    &  1&.3271(4) \\
1&.90    &  1&.308915(8) \\
1&.95   &  1&.294780(1) \\
1&.99   &  1&.280580(7) \\
\end{tabular}
\end{minipage}
\end{center}
\end{minipage}
\end{center}
\end{table}

\begin{center}
\input{paper_e4012Vsm_final.pslatex}
\end{center}

\vspace{3cm}

\begin{center}
  \input{paper_e4VsmCSW_final.pslatex}
\end{center}
\eject
\begin{center}
\input{paper_e40VsM0_final.pslatex}
\end{center}

\vspace{3cm}

\begin{center}
\input{paper_e4VsM0_final.pslatex}
\end{center}

\newpage
\linespread{1.1}

\end{document}